%% file: Main.tex
\documentclass[final,3p,times,twocolumn]{elsarticle}

\usepackage{graphicx}
\usepackage{amsmath}
\usepackage[]{xcolor}
\usepackage{ulem}
\usepackage{amssymb}
\usepackage{lineno}
\usepackage{caption}
\journal{Archives of Biochemistry and Biophysics: https://doi.org/10.1016/j.abb.2021.108923}

\begin{document}
\setlength{\parskip}{0pt} 

\begin{frontmatter}


\title{Fluid flow in the sarcomere}


\author[label1]{Sage A. Malingen}
\author[label2,label3]{Kaitlyn Hood}
\author[label4]{Eric Lauga}
\author[label3]{Anette Hosoi}
\author[label1]{Thomas L. Daniel}

\address[label1]{Department of Biology, University of Washington; Seattle, WA 98195, United States}
\address[label2]{Department of Mathematics, Purdue University, West Lafayette, IN 47906, United States}
\address[label3]{Department of Mechanical Engineering, Massachusetts Institute of Technology, Cambridge, MA 02138, United States}
\address[label4]{Department of Applied Mathematics and Theoretical Physics, University of Cambridge, Cambridge CB3 0WA, United Kingdom}

\input{Abstract.tex}


\end{frontmatter}


\input{Intro.tex}
\input{Results_n_Discussion.tex}
\input{Conclusion}

\input{Methods}

\input{Acknowledgements}
\input{DataAccessibility}


\bibliographystyle{elsarticle-num-names}
\bibliography{Bibliography.bib}
\input{Supplement}
\end{document}

%% file: Abstract.tex
\begin{abstract}
A highly organized and densely packed lattice of molecular machinery within the sarcomeres of muscle cells powers contraction. 
Although many of the proteins that drive contraction have been studied extensively, the mechanical impact of fluid shearing within the lattice of molecular machinery has received minimal attention. 
It was recently proposed that fluid flow augments substrate transport in the sarcomere, however, this analysis used analytical models of fluid flow in the molecular machinery that could not capture its full complexity. 
By building a finite element model of the sarcomere, we estimate the explicit flow field, and contrast it with analytical models. 
Our results demonstrate that viscous drag forces on sliding filaments are surprisingly small in contrast to the forces generated by single myosin molecular motors. 
This model also indicates that the energetic cost of fluid flow through viscous shearing with lattice proteins is likely minimal.
The model also highlights a steep velocity gradient between sliding filaments and demonstrates that the maximal radial fluid velocity occurs near the tips of the filaments. To our knowledge, this is the first computational analysis of fluid flow within the highly structured sarcomere. 
\end{abstract}

%% file: Intro.tex
\begin{figure}[ht]
    \centering
    \includegraphics[scale = .75]{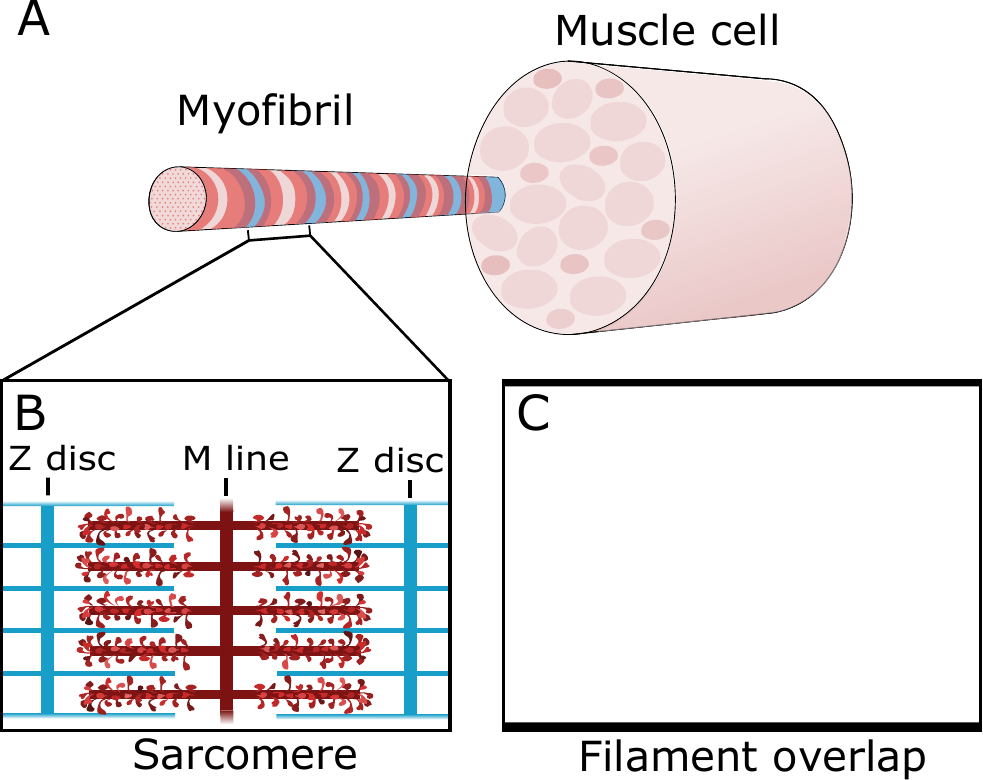}
    \caption{\textbf{{A schematic illustrating the intricate organization of muscle cells.}} A) Individual muscle cells contain many bundles of contractile machinery, the myofibrils. While not membrane bound, the myofibrils contain densely packed contractile machinery, excluding larger organelles like the sarcoplasmic reticulum and mitochondria, which occupy the space between them. B) Each myofibril is composed of sarcomeres connected in series which, in the case of skeletal muscle, run the length of the muscle cell. The sarcomere is composed of interdigitating thick and thin filament arrays which are mirrored about the m line. Molecular motors branching off of the thick filament bind to the thin filaments and pull them towards the m line, resulting in net muscle shortening. The z discs structurally anchor the thin filaments, and since they have a tightly woven structure we modeled them as impermeable boundaries. C) The thick and thin filaments form a hexagonally packed lattice. The relative ratio of thick to thin filaments varies across species and muscle types.}
    \label{musclestructure}
\end{figure}

Muscle contraction is enacted by nanometer-scale molecular machinery housed in highly organized sarcomeres (the fundamental contractile units of the muscle cell), which are connected in series running from one end of the cell to the other. 
Each sarcomere is composed of arrays of interdigitating thick and thin filaments centered on the m line (Fig \ref{musclestructure}) \cite{craig2004molecular}.
Thick filaments anchor the myosin molecular motors that power contraction when they bind to adjacent actin containing thin filaments.
Myosin binding is triggered by activation of the muscle cell, and subsequent calcium regulation of thin filament binding sites \cite{gordon2000regulation}.
Powered by the hydrolysis of adenosine triphosphate (ATP) \cite{oster2002brownian}, myosin molecular motors pull the thin filaments past the thick filaments, resulting in a net shortening of the sarcomere \cite{squire1997architecture}.
Sarcomere function relies on a panoply of proteins beyond actin and myosin (see, for instance \cite{lange2006z}), and sarcomeric proteins turnover as damaged components are removed and new proteins are incorporated \cite{willis2009build}. Calcium signalling, ATP requirements and protein turnover demand exchange between the interior of the filament lattice and the intracellular environment.
Importantly, the tightly packed lattice is percolated by fluid. And, since muscle cells are fluid filled, contraction necessitates movement of that fluid relative to the matrix of proteins. 
Although the mechanics and regulatory processes associated with muscle proteins themselves have been studied extensively \cite{Millman, squire1997architecture}, it is largely unknown how the flow of cytoplasm within the contractile lattice and the incumbent fluid forces impact function.

Two fundamental issues arise when considering the fluid environment surrounding the lattice of contractile filaments.
First, the sliding motions of filaments relative to one another will necessarily yield fluid dynamic forces.
Second, as the sarcomere shortens it is possible that there are changes in the lattice volume within the cell, with fluid moving out of the lattice as the sarcomere shortens. 
While intracellular flows are increasingly seen as drivers of general cell function \cite{mogilner2018intracellular} and fluid flow has been cited as a limitation for the rate of cellular deformation \cite{moeendarbary2013cytoplasm}, these issues have yet to be comprehensively explored in the sarcomeres of muscle cells, cellular machinery capable of kilohertz scale contraction frequencies \cite{sotavalta1953recordings}.

\begin{figure}[ht]
    \centering
    \includegraphics[scale = .74]{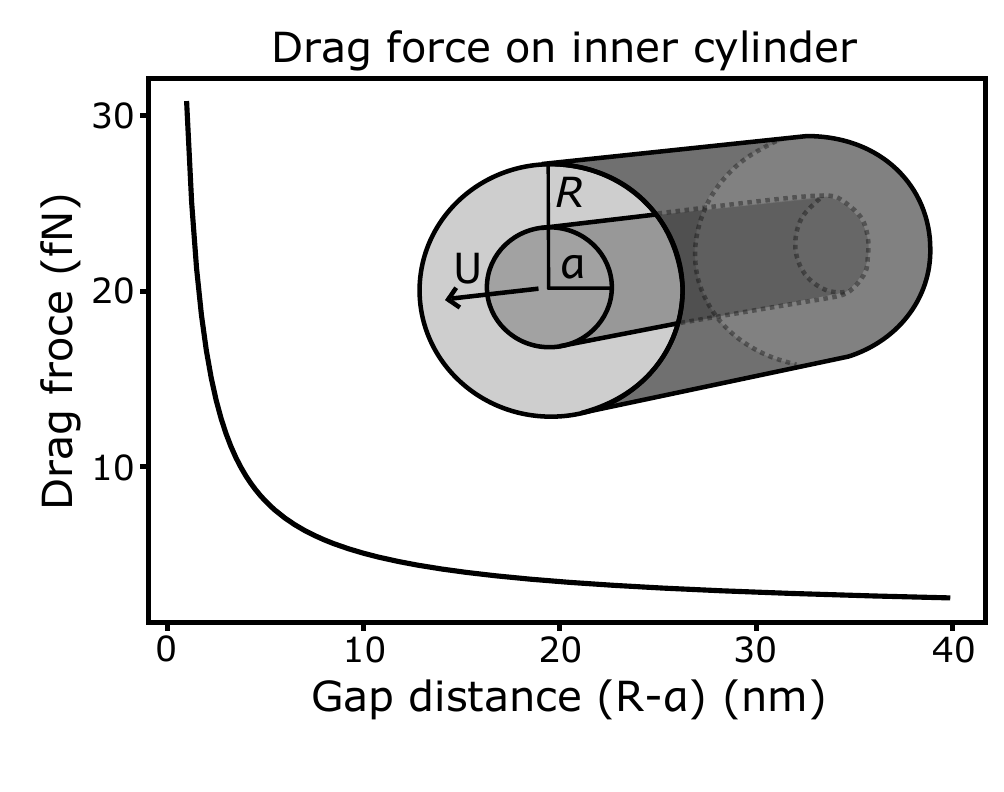}
    \caption{\textbf{The drag force on a cylinder sliding through a larger diameter cylinder depends on the gap distance between the cylinders.} Here we assumed that the fluid between the cylinders had the viscosity of water ($8.9\text{E-}4 \  \text{Pa}\cdot\text{s}$), and we prescribed a cylinder length of $1,000 \ \text{nm}$ and sliding velocity of $1,000 \ \text{nm/s}$.}
    \label{ConcentricCylinderDrag}
\end{figure}

The viscous drag forces exerted on sliding filaments in the lattice was first estimated by Huxley \cite{huxley1980reflections} by modeling thick and thin filament interactions as a slender cylinder sliding axially within a larger diameter cylinder (Fig \ref{ConcentricCylinderDrag}). He estimated the hydrodynamic force to be on the order of tens of femtoNewtons  using the estimate: 
\begin{equation}
\label{ConcentricCylinderDragEquation}
\text{Viscous drag} = -\frac{2\pi \mu U l}{\ln(a/R)}
\end{equation}
where $\mu$ is the viscosity of the fluid, $U$ is the relative velocity of the cylinders, $l$ is the length of overlap between the cylinders, $a$ is the radius of the smaller cylinder and $R$ is the radius of the larger cylinder. 
This estimate of drag is derived from the Stokes equation, continuity, and the assumption of pure axial flow. Notably, there is a typographical error in the original equation for viscous drag in \cite{huxley1980reflections}. 

Although the drag forces on individual filaments are estimated to be small, increasing the viscosity of the fluid results in decreased shortening velocity. 
While drag forces increase linearly with viscosity, increased fluid viscosity primarily slows sarcomere shortening by hampering the diffusion of molecular motors to prospective binding sites \cite{chase2000viscosity, chase1998effect}.
Additionally, the transport of calcium and ATP from the sarcoplasmic reticulum and mitochondria, respectively, to the interior of the myofilament lattice is mediated by the cytoplasm.
Despite its importance for transport and mechanical drag forces, the viscosity of the cytoplasm within the sarcomere and the relevant occlusion of the space between individual filaments remains largely unknown. 
Experimentally, both radioactive labeling \cite{kushmerick1969ionic} and fluorescence recovery after photobleaching (FRAP) experiments \cite{arrio2000translational, arrio1996diffusion} have demonstrated  that diffusion is limited in muscle cells as a function of particle size due to crowding and screening.
The constraint of diffusion time may be especially important for muscles that contract at high frequencies where substrate demands may be particularly high. 
However diffusion is not the only process mediating substrate transport.
Since the lattice of contractile machinery is not constrained to a constant volume \cite{bagni1994lattice, irving2000vivo, cecchi1990detection}, bulk fluid flow in addition to the flow driven by filament shearing must be coupled to diffusion. 
Using the diffusion, reaction, advection equation paired with analytical models of fluid flow in sarcomeres Cass \textit{et al.} demonstrated that, during cyclic contraction, bulk fluid movement augments substrate transport \cite{cass2019mechanism} over diffusion alone.
We have expanded analyses of flow in the sarcomere by building a finite element model that captures the geometry of the sarcomere, and have contrasted the resulting flow field with these analytical models.

This model examines the interactions of flows and forces for an array of sliding filaments.
This spatially explicit model also indicates that the analytical models applied in conjunction with the diffusion reaction advection equation in \cite{cass2019mechanism} are good approximations of fluid flow in many sarcomeres. 
While this model does not account for the multiscale complexity of myriad interacting sarcomeres and exterior organelles within a muscle cell, it provides a first glimpse into the understudied problem of intrasarcomeric flows.
Our results corroborate the prediction that fluid flow may impose minimal energetic consequences, while potentially augmenting substrate transport.

%% file: Results_n_Discussion.tex
\section*{Results and discussion}
We have contrasted a finite element model of fluid flow in the sarcomere with two analytical models: the first is derived from kinematic constraints while the second follows from  Darcy's law in which flow is proportional to the pressure gradient.
Both analytical models were developed fully by Cass \textit{et al} in an exploration of flow-mediated substrate transport in sarcomeres \cite{cass2019mechanism}.
By creating a computational model which explicitly captures fluid flow around filaments we have explored the effect of varying the sarcomere length (\textit{i.e} the filament overlap) and the drag forces on filaments as a function of their diameter.
Our results uniquely reveal how filaments shape the fluid flow field.

\subsection*{Key assumptions}
Due to the small spatial scale of the system, inertial forces are assumed to be negligible and the system is approximated by the Stokes equations.
Since the Stokes equations are linear, flow is reversible between contraction and elongation of a sarcomere, provided the sequence of structural changes occurring in contraction is mirrored during lengthening.
While the motions of molecular motors and changes in lattice spacing are not reversible in naturally functioning muscle, we do not expect any major differences for filament drag forces between shortening and lengthening. However, as far as the precise nature of the flow field, and its effects on transport and molecular motors, we cannot say for certain until more detailed models can be constructed

Despite the small length scale, we have assumed that the fluid is continuous and incompressible in each model (see Methods: Conceptual underpinnings).

\subsection*{Kinematic model}
Using the boundary and symmetry conditions inferred from the sarcomere's kinematics and geometry, this model provides an admissable solution to flow in the half-sarcomere.
While uniqueness is not guaranteed, these solutions for radial and axial fluid velocity ($u_r \text{ and } u_z$ respectively) satisfy the equation of continuity: 
 \begin{subequations}
 \label{FF1}
 \begin{align}
        u_r(r, z) = - U \frac{3}{4}\frac{r}{L} \bigg[ 1 - \Big(\frac{z}{L}\Big)^2 \bigg] \\
        u_z(z) = -U \frac{3}{2} \frac{z}{L}\bigg[ 1 - \frac{1}{3}\Big(\frac{z}{L}\Big)^2 \bigg]
\end{align}
\end{subequations}
where $r$ is the radial coordinate, $z$ is the axial coordinate, $U$ is the instantaneous shortening velocity of the half sarcomere, and $L$ is the axial length of the half sarcomere.
In contrast to the Darcy based model below, this kinematic model captures the physical constraint of the no-slip condition on the z disc.
However this model does not account for the structure of the myofilament lattice, which can be roughly characterized by the lattice's axial and radial permeabilities.

\subsection*{Darcy based analytical solution}
Since the sarcomere is a densely packed, anisotropic space, fluid flow may be better approximated using Darcy's law.
Although the sarcomere can be considerd to have three regions (a region with only thick filaments, a region of filament overlap and a region of thin filaments only), here we have prescribed a uniform axial permeability. 
One of the features of this model is that radial and axial permeabilities need not be equal.
This is particularly relevant for the myofilament lattice since fluid may flow more readily in the channels between filaments than across them. 

Axial flow can be described as plug flow combined with pressure-dependent Darcy flow:
\begin{equation}
\label{plug}
{u}_{{z}}({z}) = - \frac{{U}}{2} - \frac{{k}_{{l}}}{\mu}\frac{\text{d}p}{\text{d}z}
\end{equation}
where ${k}_{{l}}$ is the coefficient of axial permeability, $\mu$ denotes the fluid's viscosity and $p$ denotes pressure.
The radial component of velocity also depends on the axially varying pressure and the lattice's radial permeability, ${k}_{{r}}$.
The radial velocity is found by applying the conservation of mass, which balances radial and axial flow rates with boundary conditions: 
\begin{equation}
    {u}_{{r}}({r,z}) = \frac{{r}{k}_{r}}{{R}^{2}\mu}{p}({z}) 
\end{equation}
where ${k}_{r}$ is the coefficient of radial permeability and $R$ is the sarcomere's radius.
The pressure in the system is then derived as (see Methods):  
\begin{equation}
\label{PressureInOverlap}
{p}({z}) = \frac{\mu {UR}}{2{k}_{{l}} \alpha \sinh(\alpha {L}/2{R})}\cosh\left(\alpha \frac{{z}-{L}/2}{{R}}\right),
\ \ \alpha^{2} = \frac{2{k}_{{r}}}{{k}_{{l}}}.
\end{equation}

One advantage of this model is that the influence of radial and axial permeability can be investigated numerically by varying $\alpha$. 
Interestingly, axial and radial permeabilities may vary across muscles since, for instance, many invertebrates have different lattice packing ratios than vertebrates \cite{shimomura2016beetle}.
A limitation of this model is that it does not obey the no-slip condition at the z disc.

\begin{figure}[t]
\includegraphics[scale=.75]{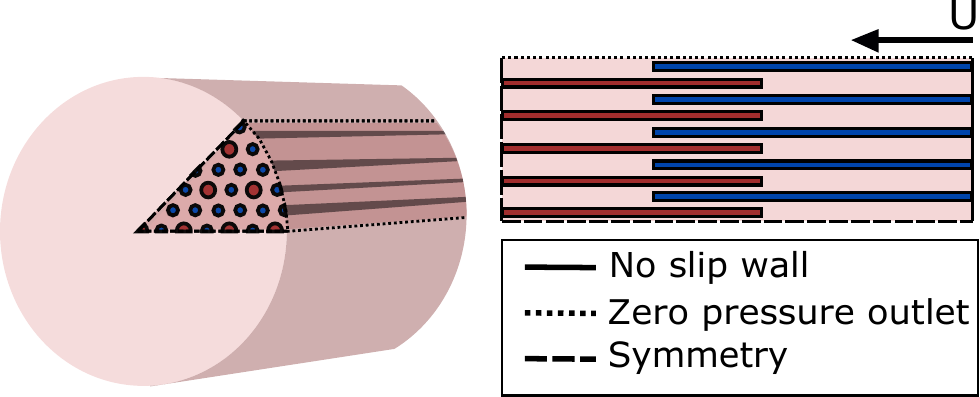}
\caption{\textbf{A schematic illustrating the geometry and boundary conditions of the finite element model.} We use symmetry to first reduce the system to a half sarcomere, and further consider a 1/8th wedge of the sarcomere. The sides of the wedge were prescribed a mirroring boundary condition while the outer edge was prescribed as a zero pressure outlet. Due to symmetry, the boundary condition at the m line is mirroring, while the boundary condition at the z disc is a no-slip moving wall sliding towards the m line at a prescribed velocity U = 1,000 nm/s. The surfaces of the thick and thin filaments also have a no-slip boundary condition, however the thick filaments are stationary, while the thin filaments move along with the z disc at a prescribed velocity of U = 1,000 nm/s.}
\label{SarcomereSchematic}
\end{figure}
\subsection*{Numerical finite element solution}
We numerically solved for the fluid flow in a simplified sarcomere-like geometry using COMSOL multiphysics' Creeping Flow interface. 
In order to reduce the computational load, which is perhaps the largest drawback of this method, we used radial symmetry and solved the flow in an axial wedge of the sarcomere (Fig \ref{SarcomereSchematic}).
We modelled the thick and thin filaments as rods and applied z no-slip boundary condition (fluid velocity at the surface is the same as the velocity of the rods).
The axial velocity at the m line was zero (by symmetry) and the velocity at the z disc was set to $-U$ in the axial component and $0$ in the radial component. 
We chose a no-slip boundary condition for the z disc since the structure is even more densely packed than the rest of the sarcomere. 
Additionally, because the motion of two adjoining sarcomeres is symmetric there is a mirroring boundary condition at the center of the z disc such that fluid cannot axially cross between two neighboring sarcomeres.
In most muscles the z disc is a narrow structure, with a notable exception found in the sonication muscle of the plainfin midshipman fish \cite{burgoyne2019three}, so in our model we have treated the z disc as a boundary.
The thin filaments moved along with the z disc, while the thick filaments were prescribed to remain stationary with the m line. 
Fluid was allowed to leave the model at the boundary by a constant pressure outlet. 

\begin{figure*}[ht]
\includegraphics[scale = .77]{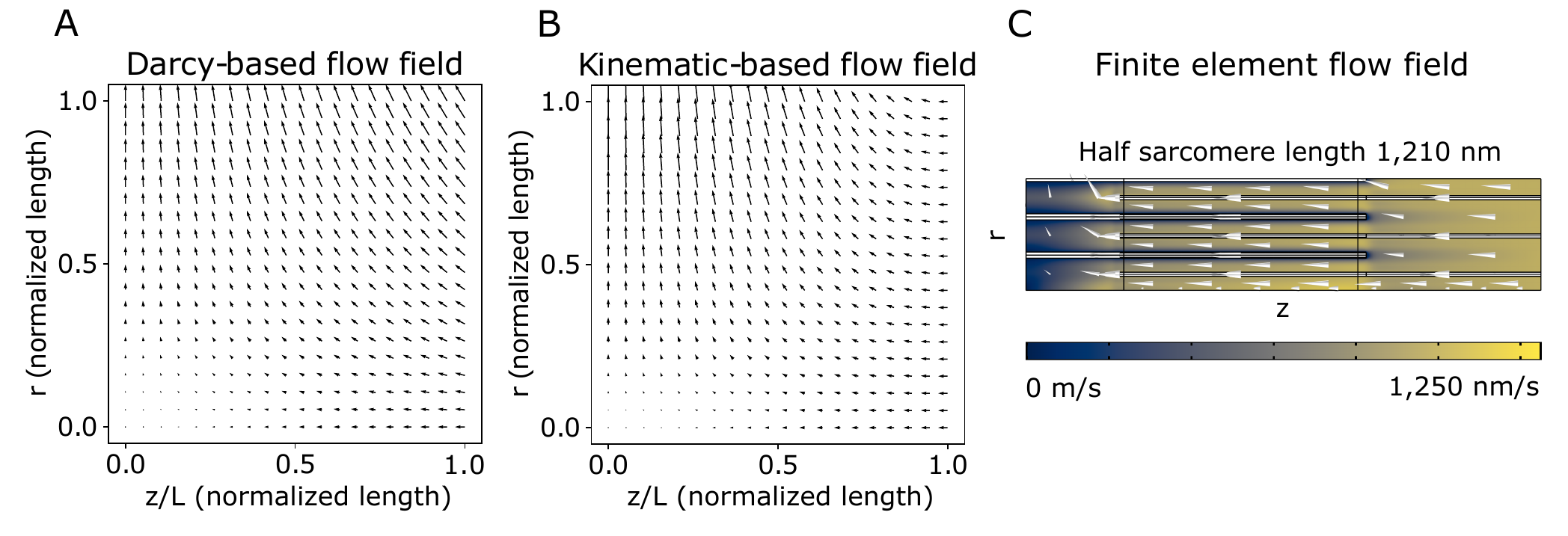}
\caption{\textbf{A comparison of the flow fields predicted by each model, where in each plot arrows denote the predicted fluid velocity.} In both panels A) and B) the vertical axis represents the normalized radial distance from the center of the sarcomere (0) to the exterior of the sarcomere (1), while the horizontal axis corresponds to the normalized axial distance from the m line (0) to the z disc (1). In panel C) the aspect ratio of the model is visually preserved. A) The Darcy-based analytical fluid flow model shows that the peak radial flow occurs at the m line, although there is significant radial flow across the sarcomere's edge, violating the no-slip condition at the z disc. B) In contrast, the analytical flow field derived from the kinematic boundary conditions of the sarcomere meets the necessary boundary conditions. It also shows a peak radial outflow at the m line and a region of stagnation at the center of the sarcomere (r = 0 and z = 0). C) In the finite element model, the velocity arrows overlay a heat map which indicates velocity magnitude. The finite element solution allows the explicit inclusion of filaments in the system, which stream the flow. The peak radial flow occurs at the tips of the thick and thin filament arrays, and there is fluid shearing between the filaments. In contrast to the analytical solutions, the fluid largely stagnates at the m line unless the thin filaments are nearly pushed up against it.}
\label{ModelCompare}
\end{figure*}

\begin{figure*}
\includegraphics[scale=.8]{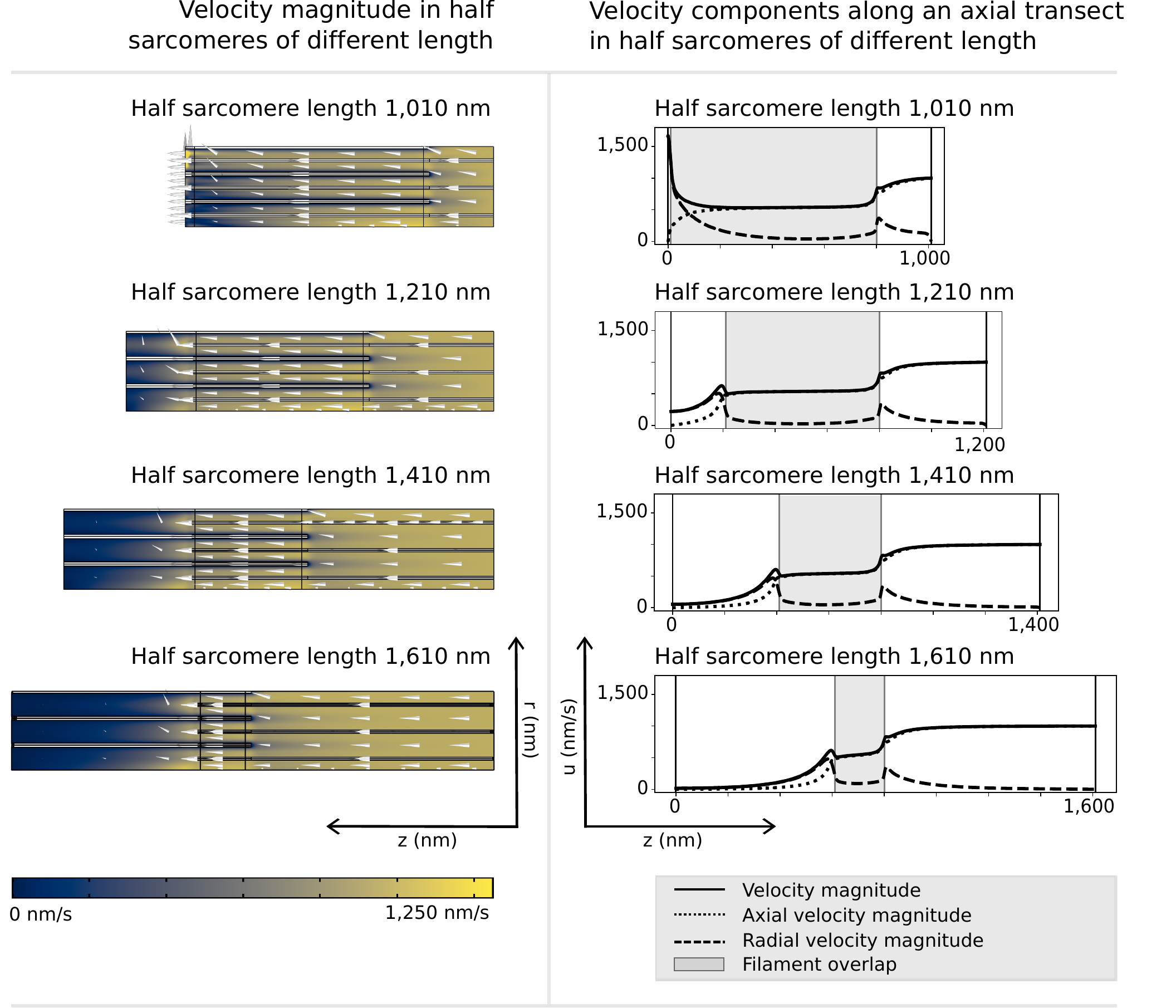}
\caption{\textbf{{The fluid flow field depends on the length of the half sarcomere as the amount of filament overlap changes.}} The velocity magnitude is shown as a heat map. Speeds over 1,250 nm/s only occurred in a small region near the m line of the 1,010 nm geometry, and these were cut off and forced to appear as 1,250 nm/s so that the finer structure of the flow fields could be compared across simulations.
The velocity along an axial transect near the border of the geometry is shown. 
The velocity is largest at the tips of filaments. 
Lattice spacing was held constant across all half sarcomere lengths.}
\label{VelPanel}
\end{figure*}

\subsection*{Comparison of models}
Each model has its advantages. 
The analytical models are computationally efficient and can be combined with other analytical models, however they neglect important structural details. 
The kinematic model accurately captures boundary conditions, but is unable to account for the internal structure of the lattice that may effect the patterns of flow.
The Darcy based model accounts for the axial and radial permeability of the lattice, but it fails to meet the no-slip condition on the z disc. 
Qualitatively these analytical models still provide similar flow fields: the fluid is ejected at the m line and axial and radial flow velocities are zero at the very center of the sarcomere due to symmetry (Fig \ref{ModelCompare}).
A key difference between them is that fluid is radially ejected along the entire length of the sarcomere's boundary in the Darcy model, while in the kinematic model the fluid has very little radial velocity near the z disc.

These qualitative results differ from the finite element model, in which the maximum radial outflow occurs at the ends of the thin filaments (Fig \ref{VelPanel}).
In fact, along an axial transect near the exterior of the filament array approximating a sarcomere the fluid has almost no radial velocity in the zone of filament overlap.
The exception to this trend is when thin filaments are nearly pressed against the m line where a mirroring boundary condition was imposed.
In this model, we observe that the fluid is ejected radially within the zone of filament overlap, as shown in the top panel of Fig \ref{VelPanel}.
Nonetheless, the analytical models may still be reasonable approximations of fluid flow in the sarcomere since many muscles operate on the ascending and plateau portions of their length tension curves \cite{ahn2018different, tu2004cardiac, burkholder2001sarcomere}, although some muscles do at times operate on their descending limb \cite{gidmark2013bite, azizi2010muscle}. 

The finite element model also draws attention to fluid shearing between individual thick and thin filaments.
How fluid shearing in this region impacts myosin molecular motor binding probability is unknown, and this model does not capture how motor molecules impact flow.

While fluid flow generally augments substrate transport to the interior of the myofibril \cite{cass2019mechanism}, the explicit nature of the flow field and its ramifications for substrate transport have not been investigated.
Although the finite element model provides valuable information about how filaments structure flow within the sarcomere, at a larger scale the spatial arrangement of organelles on the outside of the myofibril (like the sarcoplasmic reticulum, mitochondria and t-tubules) may obstruct diffusion \cite{shorten2009mathematical} and structure fluid flow.
Since these organelles act as both substrate sources and sinks their location relative to the flow field may also be important, as in other cell types \cite{mogre2020getting}. 
For instance, mitochondrial location within the cell is  constrained by both the need to acquire oxygen, and to supply myofibrils with ATP \cite{kinsey2011molecules}.
Developing a model of transport that accounts for the spatial distribution of organelles and structurally relevant flows is an exciting avenue to better determine the structural constraints of muscle cells.

\subsection*{Are viscous shearing and drag forces significant?}
Two central issues are critical in our estimate of drag forces associated with intrasarcomeric flows: (1) we do not know the viscosity of the fluid in the sarcomere and (2) there is an unknown extent of water bound to the filaments themselves. 
In addition to layers of bound water, the effective diameter of the thick filament may depend on the extension of molecular motors away from the thick filament's backbone. Here we address these issues explicitly.  

\begin{figure}
\includegraphics[scale=.78]{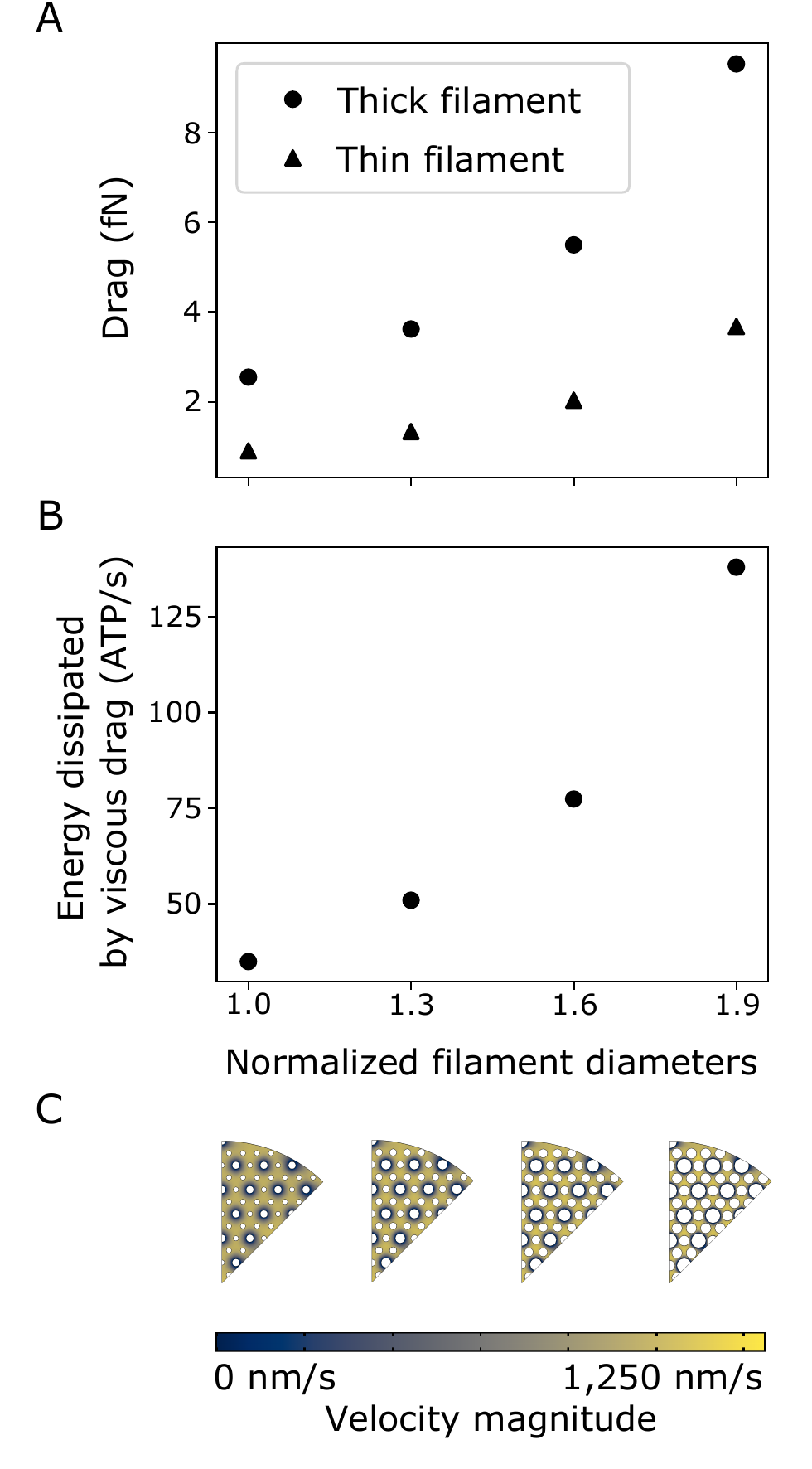}
\caption{\textbf{A finite element model demonstrates that drag forces, along with the energy dissipated by drag forces, increase as filament diameters increase in a sarcomere-like geometry.} A) Drag force increases with increasing filament diameters. Drag forces were measured using a surface integration on a thin and a thick filament near the center of the sarcomere. We multiplied the filament diameters by a scaling factor (normalized filament diameter) till their surfaces nearly touched since their effective \textit{in vivo} diameters are unknown.
B) The energy dissipated by viscous drag forces on the sliding filaments was estimated by assuming that the sarcomere is composed of 500 thick filaments and 3,000 thin filaments and shortens at 1,000 nm/s. Although more energy is dissipated by viscous drag as filament diameter is increased, the overall energetic expense is relatively small compared to the cell's overall energy demands. C) The diameter of the filaments was increased by multiplying their base diameter by a scaling factor: the normalized filament diameter. To illustrate how the geometry changes as filament diameter is increased we have shown a cross-sectional heatmap of velocity magnitude for each of the simulations.}
\label{DragForces}
\end{figure}

In the packed intra-cellular space, factors such as macromolecular crowding and bound water may  significant influence  drag forces and diffusion mediated processes \cite{luby1994physical}.
Although the depth of bound water layers around the thick and thin filaments is unknown, osmotic compression indicates that $~30\%$ of the myofilament lattice is osmotically inactive, with a protein volume of $20-25\%$, indicating that $5-10\%$ of the volume is composed of bound cytoplasm \cite{Millman}.
Additionally, the extent to which myosin motors are extended from the backbone, or held against it, depends on a host of regulatory factors \cite{powers2021sliding}.
Using a parameter sweep, we estimated how the thickness of the filaments impacted drag forces.

It is also vitally important to note that the sarcomere is packed with myriad regulatory and structural proteins that we have not included in our model. 
How their precise structures, locations and motions impact fluid flow and drag forces is unknown.
While increasing filament diameters accounts to some extent for the occlusion of the space, it does not account for how radial protein structures may prevent channeling of fluid between the thick and thin filaments.

Using our finite element model we estimated the drag force on a single thick filament near the center of the sarcomere, and a neighboring thin filament.
Our results confirm an early hydrodynamic estimate that drag forces are small \cite{huxley1980reflections}.
Although drag force increases with filament diameter, even when filaments were nearly touching one another viscous drag was still surprisingly small when contrasted with the picoNewton scale forces generated by individual molecular motors \cite{piazzesi2007skeletal, molloy1995movement} (Fig \ref{DragForces}).
We note that the viscosity of water is likely a lower bound for the viscosity of the cytoplasm, and drag forces scale linearly with viscosity.
However, even scaling the viscosity up to $1,000\times$ that of water the drag forces are still of the same order of magnitude as the forces generated by single myosin molecular motors \cite{linari2007stiffness}.
These results support earlier findings that demonstrated increasing fluid viscosity reduced contraction velocity by slowing cross-bridge kinetics, rather than by imposing drag forces on filaments that hampered contraction \cite{chase1998effect}.

It is reasonable to ask how fluid moves in response to the lattice’s shape change, but also how fluid flow influences the lattice’s volume.
Since the drag forces our model predicts are small when contrasted with the forces generated by cross bridges we expect that fluid flow is not the primary determinant of lattice volume changes, but instead that it flows in response to the lattice's motions.

While the forces themselves do not prohibit filament sliding, it is interesting to ask if the total energetic cost of overcoming viscous shearing is similarly negligible. 
To address this issue, we used our computed drag forces for thin and thick filaments at the center of the sarcomere along with their velocity to compute the rate of energy expenditure by viscous shearing.  
Then, for an idealized sarcomere that contains 500 thick filaments and 3,000 thin filaments and that shortens at 1,000 nm/s, we computed the lower-bound total rate of viscous energy dissipation to be 0.004 fW. 
Since this is not an intuitive unit, we can frame this number in a more biological context by observing that the energy released by ATP hydrolysis is 69 kJ/mole \cite{wackerhage1998recovery} and then converting from W to ATP molecules consumed per second.
We estimate at the lower bound that approximately 35 ATP molecules per second are consumed by viscous drag forces.
This estimate was computed using the minimum filament diameters, the viscosity of water, a 1:3 thick to thin filament packing ratio and D10 of 45 nm (corresponding to a surface to surface gap distance of ~20 nm).

Although the 1:3 packing ratio we used is common among invertebrate muscles, there is considerable taxonomic diversity in packing ratio \cite{shimomura2016beetle}.
In vertebrates, a 1:2 packing ratio is standard. 
However among invertebrate taxa there are multiple examples of muscles with many more thin filaments than either the 1:2 or 1:3 packing ratios. 
To investigate how drag forces depend on packing ratio, we computed the fluid flow for a sarcomere with a packing ratio of 1:5 (which is approximately the packing ratio of cockroach femoral muscle \cite{hagopian1966myofilament}, although it has also been estimated at a 1:6 packing ratio \cite{jahromi1969structural}). 
With a smaller thick to thin filament ratio, the thick filament drag force magnitude was slightly larger than that of the corresponding model with a 1:3 packing ratio (3.2 fN to 2.6 fN respectively), and a slightly smaller thin filament drag force (0.6 fN to 0.9 fN respectively). 

Although these drag estimates remain small compared to the forces generated by individual molecular motors, viscous drag forces could have broader ramifications, such as pushing molecular motors into a different position and altering their binding probability.
Since contraction of the micron-scale sarcomere is enabled by the cyclic action of nanometer-scale molecular motors working in concert, small reorientations could have a large effect. 
Additionally, the effective diameter of the filaments given bound water and the presence of molecular motors projecting from the backbone has received minimal attention. 

\subsection*{Future horizons}
Our analytical and numerical approaches are based on a continuum mechanics model of flow in the myofilament lattice, revealing that the geometry of the lattice shapes the fluid flow field, and influences forces.
Each approach stems from finding an approximate solution to the Navier-Stokes equation\textcolor{red}{s} for a geometry and motion similar to those observed in nature.
The continuum approach assumes that individual molecular interactions can be averaged since the Knudsen number is
\begin{equation}
\label{Knudsen-number}
\text{Kn} = \frac{\lambda}{L} \approx \frac{0.3 \text{nm}}{20 \text{nm}} = 0.015
\end{equation}
given a mean free path ($\lambda$) of liquid water of approximately 0.3 nm \cite{padding2006hydrodynamic} and a representative length scale for the geometry ($L$) of 20 nm. 
However, sub-sarcomeric fluid mechanics is close to the spatial scale at which the continuum hypothesis breaks down (Kn $\approx$ 0.1), and we note that future bulk hydrodynamic modeling may need to be coupled with methods that account for non-continuum effects. 
This may be especially true for investigating the ramifications of fluid flow around molecular motors and regulatory proteins.
Approaches that may be effective include Direct Simulation Monte Carlo (DSMC, which takes a probabilistic view of particle location based on the Boltzmann equation) \cite{alexander1997direct}, fluctuating hydrodynamics (which blends stochasticity with the deterministic Navier-Stokes equation\textcolor{red}{s} to approximate particle effects at a bulk scale \cite{chaudhri2014modeling}) or molecular dynamics (MD, which, although computationally expensive and computationally unrealistic at the sarcomere scale, explicitly account for particle collisions) \cite{nie2004continuum}.
Due to the range of time and spatial scales represented, estimating the dynamics of the cytoplasm during muscle contraction will likely require techniques that coarse-grain individual particle motions, coupling large scale hydrodynamics with Brownian motion \cite{padding2006hydrodynamic}.

\begin{figure}
    \centering
    \includegraphics[scale = .72]{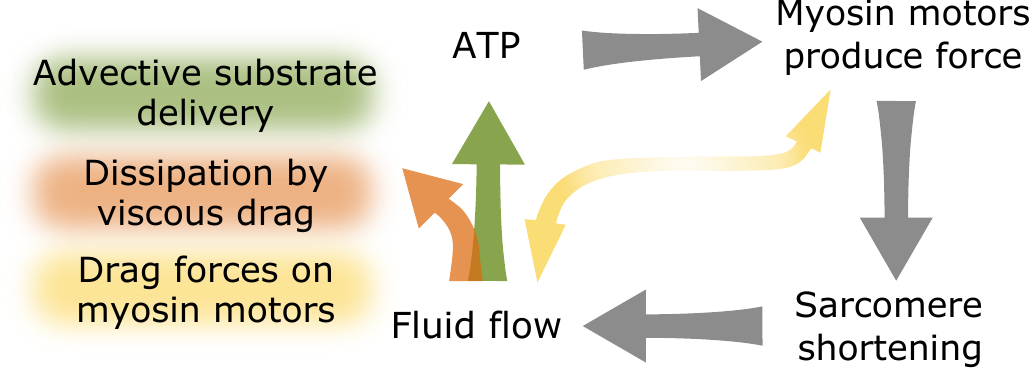}
    \caption{\textbf{Fluid flow in the sarcomere may impact function through multiple mechanisms.} ATP powers the shape change of the myosin molecular motors, which ultimately results in sarcomere shortening. Due to the shortening of the sarcomere, there is a flow of fluid, potentially in and out of the lattice. The movement of fluid could impact function by augmenting substrate delivery to the interior of the densely packed lattice. The shearing of filaments with the fluid will also result in the dissipation of energy by viscous drag. The drag force of the fluid on molecular motors could bias their positions and alter their binding probabilities. Whether this would have a positive or negative impact on contractility is unknown, and would likely depend on the characteristics of the flow field. }
    \label{flowcycle}
\end{figure}

While in this study we highlighted the flow of fluid within the sarcomere, fluid flow across organelles, from the interacting flow fields induced by neighboring sarcomeres to the movement of fluid around the sarcoplasmic reticulum and mitochondria, may have functional implications that depend explicitly on cell geometry.
Thus the scales over which models must couple may range from nanometers (for instance, as drag forces may alter the dynamics of molecular motors and other regulatory proteins, and spatially explicit fluid flow may reveal directed substrate transport) to millimeters (as fluid flow mediates substrate transport within whole cells). 
To address these issues, future computational efforts can build upon the spatially explicit models of flow that we have developed here.  
Integrating from molecular to cell scales is an exciting horizon for cell biologists necessitating an understanding Brownian dynamics, colloidal dynamics and continuum mechanics \cite{maheshwari2019colloidal}.
Striated muscle boasts a uniquely organized and well-studied system to build models that integrate across these scales. 

%% file: Conclusion.tex
\section*{Conclusion}
We contrasted two analytical models of fluid flow in the sarcomere with a finite element model created with COMSOL to characterize flow fields and forces in a sarcomere-like geometry. 
The fluid flow field is significantly impacted by the presence of filaments occluding the space, with the regions of largest velocity magnitude near the z disc, and maximum radial outflow at the tips of the filaments.  When the tips are at the m line, radial flow fields are more similar to the analytical results, and the maximum radial velocity occurs at the m line.  Interestingly, such flows correspond to conditions of maximum filament overlap which is physiologically quite common.  

Both drag forces and diffusion are influenced by a number of hard-to-characterize parameters: the viscosity of the cytoplasm, the degree of molecular crowding (and the size of the molecules crowding the space), layers of bound water and electrostatic interactions.
Despite these uncertainties, we suggest that the energetic cost induced by viscous dissipation is small compared to the cell's overall energy use.
However it is important to note that we assumed a steady sliding velocity, rather than sliding induced by impulsive forces. The latter could drastically increase viscous drag forces on sliding filaments \cite{elliott2001muscle}.
Overall, the generally tiny viscous forces suggest that advective flow may be an energetically inexpensive mechanism that augments substrate transport, and fluid flow could bias cross bridge binding with as yet unknown effects.

%% file: Methods.tex
\section*{Methods}
\subsection*{Conceptual underpinnings}
We estimated fluid flow in the sarcomere with analytical models and a finite element based approach using COMSOL.
Each ultimately derives from finding an approximate solution for the Navier-Stokes equation\textcolor{red}{s} for a geometry and motion similar to those observed experimentally.
The Navier-Stokes equations (equation \ref{Navier-Stokes}) along with the equation of continuity fully describe the movement of a fluid \cite{bird2002transport}:
\begin{equation}
\label{Navier-Stokes}
 \rho \left(\frac{\partial\boldsymbol{u}}{\partial t} + \boldsymbol{u}\cdot\nabla\boldsymbol{u}\right)= - \nabla p + \mu \nabla^{2} \boldsymbol{u} + \boldsymbol{f}
\end{equation}
where $\rho$ is the mass density of the fluid, $t$ denotes time, $\mathbf{u}$ is a vector field describing the fluid's velocity, $p$ is a scalar field accounting for pressure, $\mu$ is the fluid's viscosity and $\boldsymbol{f}$ captures external body forces (such as gravity).
Positing that the sarcoplasm is incompressible and that the flow is steady, the continuity equation expressed in cylindrical coordinates is: 
\begin{equation} 
\label{Continuity_full}
\frac{1}{r}\frac{\partial}{\partial r}(ru_{r}) + \frac{1}{r}\frac{\partial u_{\theta}}{\partial \theta} + \frac{\partial u_{z}}{\partial z} = 0.
\end{equation}
Considering axisymmetric flow, the continuity equation reduces to: 
\begin{equation} 
\label{Continuity_axisymmetric}
\frac{1}{r}\frac{\partial}{\partial r}(ru_{r}) + \frac{\partial u_{z}}{\partial z} = 0.
\end{equation}
 
Next we consider the Reynolds number, which is the ratio of inertial to viscous stresses in the system.
Given a characteristic length scale of $3 \ \mu \text{m} = 3\times10^{-6} \ \text{m}$ (which is a common sarcomere length), velocity of $1 \ \mu \text{m}/\text{s} = 1\times10^{-6} \ \text{m}/\text{s}$, viscosity of $8.9\times10^-4 \ \text{Pa}\cdot\text{s}$ and density of 997 $\text{kg}/\text{m}^3$ the Reynolds number is: 
\begin{equation}
\begin{aligned}
    Re &= \frac{{DU}\rho}{\mu} \\
       &= \frac{1\times10^{-6} \cdot 1\times10^{-6} \cdot 997}{8.9\times10^{-4}} \\
       &\approx 3\times10^{-6} \ll 1
\end{aligned}
\end{equation}

A Reynolds number much less than one means that viscous forces predominately determine the behavior of the system, and so inertial forces can be neglected. 
Therefore we can simplify the system of governing equations, yielding the linear Stokes equations:
\begin{equation}
\label{Stokes}
 \nabla p = \mu \nabla^{2}\boldsymbol{u} + \boldsymbol{f}
\end{equation}
Since the Stokes equations are linear, the flow elicited by a sequence of geometry changes is precisely reversible if the sequence of geometric changes is performed in exact reverse.

These governing equations undergird the modeling methods we used.
Ultimately, each of the modeling methods we used searches for a possible solution to these equations subject to the boundary conditions posed by the geometry and motion of the sarcomere.
Here we provide a brief explanation of the Darcy-based and kinematic-based fluid flow models which are fully developed by Cass \textit{et al} \cite{cass2019mechanism}.

\subsection*{Development of Darcy based model}
This model combines plug flow and Darcy based flow, which allows the variation of permeability. 
The half sarcomere can be considered to have three regions (thick filaments only, filament overlap and thin filaments only) that change in proportion with shortening.
This model simplifies the problem by considering only a single region of overlapping filaments and approximating flow as a combination of plug flow and Darcy flow due to the motion of the z disc and the presence of filaments occluding the space.
Plug flow varies only as a function of axial location and decreases as a function of the change in pressure and the resistance of the lattice to fluid flow:
\begin{equation}
{u}_{z} (\text{z}) = - \frac{U}{2} - \frac{k_{l}}{\mu}\frac{\text{d}p}{\text{d}z},
\label{eq12}
\end{equation}
where $k_l$ is the longitudinal permeability of the fibers and is proportional to the inter-fiber distance squared ($k_l \sim \delta^{2}$). 
In our case, the inter-fiber distance $\delta = \frac{D10}{\sqrt{3}}$.

To derive the pressure term in Eq \ref{eq12}, we note that mass conservation demands the radial and axial volume flows must balance
\begin{equation}
    \frac{\text{d}}{\text{dz}}(\pi \text{R}^{2}{u}_{z}) = -2 \pi \text{R}  {u}_{\text{r}}(\text{R,z}).
\end{equation}
We then invoke a Darcy relationship at the edge of the half sarcomere ($r = R$) where ${k}_{r}$ denotes the radial permeability of the fiber bundle:
\begin{equation}
\label{radialDarcy}
{u}_{r}(R,z) = \frac{{k}_{r}}{R\mu}p(z). 
\end{equation}
By pairing these we come to an equation describing the pressure:
\begin{equation}
    \frac{\text{d}^{2}p}{\text{d}{z}^{2}} = \frac{\alpha^{2}}{R^{2}}p(z), \ \ \alpha^{2} = \frac{2k_{r}}{k_{l}}.
\end{equation}
Since this is a second order linear homogeneous differential equation, the solution takes the form of a superposition of exponentials, which can be expressed as the sum of a hyperbolic sine and hyperbolic cosine. 
While there are two roots of the equation, it can be simplified since $k_{l}, k_{r}$ and $R$ are all positive.
Hence the general solution for pressure is:
\begin{equation}
    p(z) = {p}_{1} \text{cosh} \left(\alpha \frac{z}{R}\right) + {p}_{1} \text{sinh} \left(\alpha \frac{z}{R}\right).
\end{equation}
Given the boundary conditions for pressure:
\begin{equation}
    \frac{\text{d}p}{\text{d}z}\bigg\rvert_{\text{L}} = \frac{\mu {U}}{2{k}_{{l}}}, \ \frac{\text{d}p}{\text{d}z}\bigg\rvert_{0} = -\frac{\mu {U}}{2{k}_{{l}}}
    \label{PressureBoundaries}
\end{equation}
and the fact that the solution for pressure is even with respect to the mid-point $z = L/2$, where $L$ is the half sarcomere's length, the solution can be simplified to:
\begin{equation}
p(z) = p_{1}\text{cosh}\left(\alpha \frac{z-L/2}{R}\right),
\end{equation}
where:
\begin{equation}
    p_{1} = \frac{\mu UR}{2k_{l}\alpha \text{sinh}(\alpha L/2R)}.
\end{equation}
Then the radial velocity at the sarcomere's edge ($R$) is:
\begin{equation}
    u_{r}(R,z) = \frac{U\alpha}{4}\frac{\text{cosh}\left(\alpha \frac{z-L/2}{R}\right)}{\text{sinh}\left(\alpha L/2R\right)}.
\end{equation}

The radial flow within the sarcomere can be obtained from the conservation of mass. 
By taking a local average, and assuming flow is axisymmetric, we obtain:
\begin{equation}
    \frac{\partial {r} {u}_{{r}} }{\partial r} = -{r} \frac{\text{d}u_{z}}{\text{d}z}.
\end{equation}
Assuming that this expression is regular about $r = 0$, integrating by $r$ yields:
\begin{equation}
    {u}_{r}(r,z) = -\frac{r}{2}\frac{\text{d} u_{z} }{\text{d}z}.
\end{equation}
Then since 
\begin{equation}
    \frac{\text{d}u_{z}}{\text{d}z} = -\frac{k_{l}}{\mu}\frac{\text{d}^{2}p}{ \text{dz}^{2}} = -\frac{{k}_{{l}}\alpha^{2}}{\mu R^{2}}p({z}) = -\frac{2{k}_{r}}{\mu R^{2}}p(z),
\end{equation}
we finally obtain 
\begin{equation}
 {u}_{{r}}(r,z) = \frac{{r}{k}_{r}}{R^{2}\mu}p(z) = \frac{r}{R} u_{r} (R,z).
\end{equation}

\subsection*{Kinematics based model}
The kinematic based model was postulated from the system's boundary conditions.
While uniqueness is not shown, this solution is continuous and meets the available kinematic conditions, so it is an admissible solution.
First, by the no slip condition, fluid at the z disc has zero radial velocity:

\begin{equation}
    u_{r}(r, L) = 0.
\end{equation}

The m line and the radial center of the sarcomere also define two planes of symmetry. 
At the m line there can be zero net axial shear, while along the axial center there can be no net radial shearing:
\begin{equation}
    \frac{\partial}{\partial z} u_{z}(r, 0) = 0,
\end{equation}
\begin{equation}
    \frac{\partial}{\partial z} u_{r}(0, z) = 0.
\end{equation}
In addition to the equation of continuity these conditions constrain the solution space, and they are met by the model: 
\begin{subequations}
\begin{align}
        u_r(r, z) = - U \frac{3}{4}\frac{r}{L} \bigg[ 1 - \Big(\frac{z}{L}\Big)^2 \bigg], \\
        u_z(z) = -U \frac{3}{2} \frac{z}{L}\bigg[ 1 - \frac{1}{3}\Big(\frac{z}{L}\Big)^2 \bigg].
\end{align}
\end{subequations}

\subsection*{Finite element model}
The geometry of the model was based on experimental observations. 
Although the lengths of the thick and thin filaments depend on the species and muscle type, the values we used are similar to a variety of muscles in common model organisms \cite{shimomura2016beetle, al2013three}.
The spacing of the filament lattice also varies depending on organism, muscle type, and even the operating conditions for a single muscle.
Our choice is based on \textit{Manduca sexta} during \textit{in vivo} function \cite{malingen2020vivo}.
In our model we held lattice spacing constant, although lattice spacing changes commonly occur over the course of contraction in living systems \cite{powers2021sliding}. 
Finally, the diameter of the thick and thin filaments was chosen based on atomic reconstructions of the filaments, \cite{woodhead2005atomic} and \cite{holmes1990atomic}, respectively, and electron microscopy \cite{hayes1971electron}.
The diameters of the filaments relevant for hydrodynamic interactions is unknown since layers of water may be bound to filament surfaces, and since myosin molecular motors extend from the thick filament backbone.
Therefore we chose conservative base values and later scaled these values with a dimensionless factor from 1 to 1.9.
We used a model with a 1:3 thick-to-thin filament packing ratio for the base simulations. 
We also used a separate model with a 1:5 thick-to-thin filament packing ratio since there is natural variation in packing ratios across taxa. 
We took advantage of symmetry to reduce the computational cost of the simulation, first reducing the simulation to a half sarcomere from the m line to z disc, and then further constraining it to a one-eighth wedge, resulting in one-sixteenth of a sarcomere. 
In order to reduce the computational load we limited the model's radius to be six times the lattice's spacing (the D10). 
The boundary conditions we prescribed for the model are illustrated in Fig \ref{SarcomereSchematic}, briefly, the z disc was prescribed as a no slip wall, the m line and adjoining radial faces were prescribed as a symmetry plane, the outer face along the circumference was a zero-pressure outlet and the surfaces of the filaments were prescribed a no-slip condition. 

\begin{table}[ht]
\begin{center}
\begin{tabular}{ |l| c | }
\hline
 Structure & Size (nm) \\ 
 \hline\hline
 Thin filament radius & 5 \\ 
  \hline
 Thin filament length & 1000 \\  
  \hline
 Thick filament radius & 7 \\
  \hline
 Half thick filament length & 800 \\
 \hline
 Lattice spacing (D10) & 45 \\
 
 \hline
\end{tabular}
\caption{The base values used for the model geometries are provided. Filament lengths were chosen based on commonly observed values in a variety of organisms \cite{shimomura2016beetle, al2013three}. It should be noted that these lengths correspond to only the portions of the thick and thin filaments within a single half sarcomere (z disc to m line), so we refer to a half thick filament length, and thin filament length. The spacing of the filament lattice is also variable, however our choice is commonly observed in \textit{Manduca sexta} during \textit{in vivo} function \cite{malingen2020vivo}. The diameters of the thick and thin filaments were chosen based on atomic reconstructions of the filaments, \cite{woodhead2005atomic} and \cite{holmes1990atomic}, respectively, and electron microscopy \cite{hayes1971electron}.}
\end{center}
\end{table}

\subsubsection*{Simulation details}
\subsubsection*{Creeping flow simulations}
We used a free tetrahedral mesh paired with COMSOL's option for a normal mesh element size calibrated for fluid dynamics. 
The maximum element size was 1.49E-8 m and the minimum element size was 4.44E-9 m with a maximum element growth rate of 1.15 and curvature factor of 0.6 and 0.7 resolution of narrow regions.

%% file: Acknowledgements.tex
\section*{Acknowledgements}
The authors are grateful to Jim Riley and G.M. ``Bud'' Homsy for helpful discussions about low Reynolds number fluid mechanics, and to Julie Theriot for discussing viscosity and transport in living cells.
Dave Williams graciously provided the 3D rendering of the thick and thin filaments used in figure 2. 
The Army Research Office (W911NF-14-1-0396 to TLD and AH) and the Joan and Richard Komen Endowed Chair to TLD provided support to this project, along with NIH grant P30AR074990.
SAM was funded by a Bioengineering Cardiac Training Grant from the National Institute of Biomedical Imaging and Bioengineering (T32EB1650) and a fellowship from the ARCS Foundation.
This project also received funding from the European Research Council (ERC) under the European Union's Horizon 2020 research and innovation programme (grant agreement 682754 to EL).
KTH was supported by National Science Foundation Grant DMS-1606487.

%% file: DataAccessibility.tex
\section*{Data Availability}
The Comsol models can be accessed via the Dryad digital repository at: https://doi.org/10.5061/dryad.q2bvq83jb.

%% file: Supplement.tex
\section*{Supplement}
\section*{Simulation details}
\begin{table*}[]
\begin{center}
\caption{Simulation details for each of the sarcomere length, filament diameter scaling and ratio of filament combinations we used are provided. }
\begin{tabular}{ |c| c | c | c | c | c | c | c | c | }
\hline
 Length & Scaling & Ratio & Degrees of freedom & Domain elements & Boundary elements & Edge elements \\ 
  \hline\hline
  1010 (nm) & 1 & 1:3 & 11516708 & 2586291 & 247141 & 43796 \\ 
  \hline
  1210 (nm) & 1 & 1:3 & 12763833 & 2879599 & 254068 & 43850 \\  
  \hline
  1410 (nm) & 1 & 1:3 & 14017703 & 3174545 & 261254 & 43934 \\
  \hline
  1610 (nm) & 1 & 1:3 & 14439485 & 3276152 & 259970 & 42883 \\
  \hline
  1410 (nm) & 1.3 & 1:3 & 12298304 & 2723526 & 309126 & 43762 \\
  \hline
  1410 (nm) & 1.6 & 1:3 & 8333899 & 1793453 & 281440 & 41141 \\
  \hline
  1410 (nm) & 1.9 & 1:3 & 4978909 & 1011241 & 249279 & 35628 \\
  \hline
  1410 (nm) & 1 & 1:5 & 14363061 & 3196887 & 348618 &  67284\\
  \hline
\end{tabular}
\end{center}
\end{table*}